\documentclass[12pt]{article}
\usepackage{ametsoc}

\usepackage{graphicx,natbib,subfigure}
\usepackage{color}
\usepackage{changes}
\usepackage{braket}
\usepackage{siunitx}
\usepackage{lineno}
\linenumbers 

\newcommand{\lam}{\lambda}

\newcommand{\jacob}[1]{\sqrt{\mathcal{J}^{\dagger}\mathcal{J}}}

\newcommand{\myabstract}{The selection process of proposals is a crucial component of scientific progress and innovations. Limited resources must be allocated in the most effective way to maximise advancements and the production of new knowledge, especially as it is becoming increasingly clear that technological and scientific innovation and creativity is an instrument of economic policy and social development. The traditional approach based on merit evaluation by experts has been the preferred method, but there is an issue regarding to what extent such a method can also be an instrument of effective policy.  This paper discuss some of the basic processes involved in the evaluation  and selection of proposals, indicating some criterion for an optimal solution..}
\begin{document}
%
%
\title{\textbf{\large{ The choice: evaluating and selecting scientific proposals}}}
%
%
\author{\textsc{A.Navarra}
				\thanks{\textit{Corresponding author address:} 
				Antonio Navarra, Viale Aldo Moro 44, 40129 Bologna, Italy. 
				\newline{E-mail: antonio.navarra@cmcc.it }}\\
\textit{\footnotesize{Centro Euromediterraneo sui Cambiamenti Climatici, Bologna, Italy}}\\
 \textit{\footnotesize{and}}\\
 \textit{\footnotesize{ Istituto Nazionale di Geofisica e Vulcanologia, Bologna, Italy}}
}
%
\ifthenelse{\boolean{dc}}
{
\twocolumn[
\begin{@twocolumnfalse}
\amstitle

\begin{center}
\begin{minipage}{13.0cm}
\begin{abstract}
	\myabstract
	\newline
	\begin{center}
		\rule{38mm}{0.2mm}
	\end{center}
\end{abstract}
\end{minipage}
\end{center}
\end{@twocolumnfalse}
]
}
{
\amstitle
\begin{abstract}
\myabstract
\end{abstract}
\newpage
}
%

\section{Introduction}

The selection process of  scientific activities is usually performed by review of the merit and originality of the proposal. It is generally thought that such a competitive process will guarantee the emergence of the projects with the higher potential to offer new insights and produce more innovation. However, research is a very uncertain business and the ex-ante innovation potential may not be realised at all or achieved only partially. The evaluation procedures are obviously measuring only the {\em potential} of innovation and scientific advance and therefore are by definition affected by errors. Such errors are judgemental, cultural, numerical or simply social as they result from different school of thought or various factions in the scientific community.
It is reasonable to ask if there is a "best" strategy in the shaping of a research program to yield an optimal level  of selection for the proposal that would take into account these uncertainties. 

A naive approach would simply  use only the very top projects, but this choice would result in a reduced diversification of approaches and methods, thereby increasing the risk. On the other hand, accepting all projects will guarantee the maximum innovation, but it will be wasteful of resources and morally unacceptable because there will be no incentive to produce, sound, well-based proposals.

Detailed analysis of project selection including asymmetric informations and outside choices have been performed \citep{Talia}using highly sophisticated mathematical methods, but this short paper proposes to analyse this mechanism via a  simple model of evaluation and project distribution, resulting in an understanding of the underlying mechanism. The approach allows to estimates some optimal thresholds for maximising scientific results and innovation.

\section{The model}

The various functions and distributions in this paper will be described in terms of the {\em evaluation value}, $x$, that is the score obtained via an ex-ante evaluation by a certain proposal, project or other forms of scientific documents as a result of a solicitation or a call for tender. 

The detailed evaluation procedure is not important here, but only that whatever score is used it must be a monotonic function of the implicit "value" of the proposal. This is not such  a strong restriction since any indicator of value should realise a consistent ranking of proposals. 

The density of proposals as a function of the score will be denoted by $p(x)$ so that the number of proposal up to a certain score $\lam$ will be given by
\begin{equation}
N(\lam) = \frac{1}{N_1}\int_0^\lam p(x) \, dx
\end{equation}
normalized by the total number of proposals,
$$
N_1 = \int_0^1 p(x) \, dx.
$$

\begin{figure}[h!]
\centering
\includegraphics[width=\columnwidth,clip,trim= 0 200 0 200]{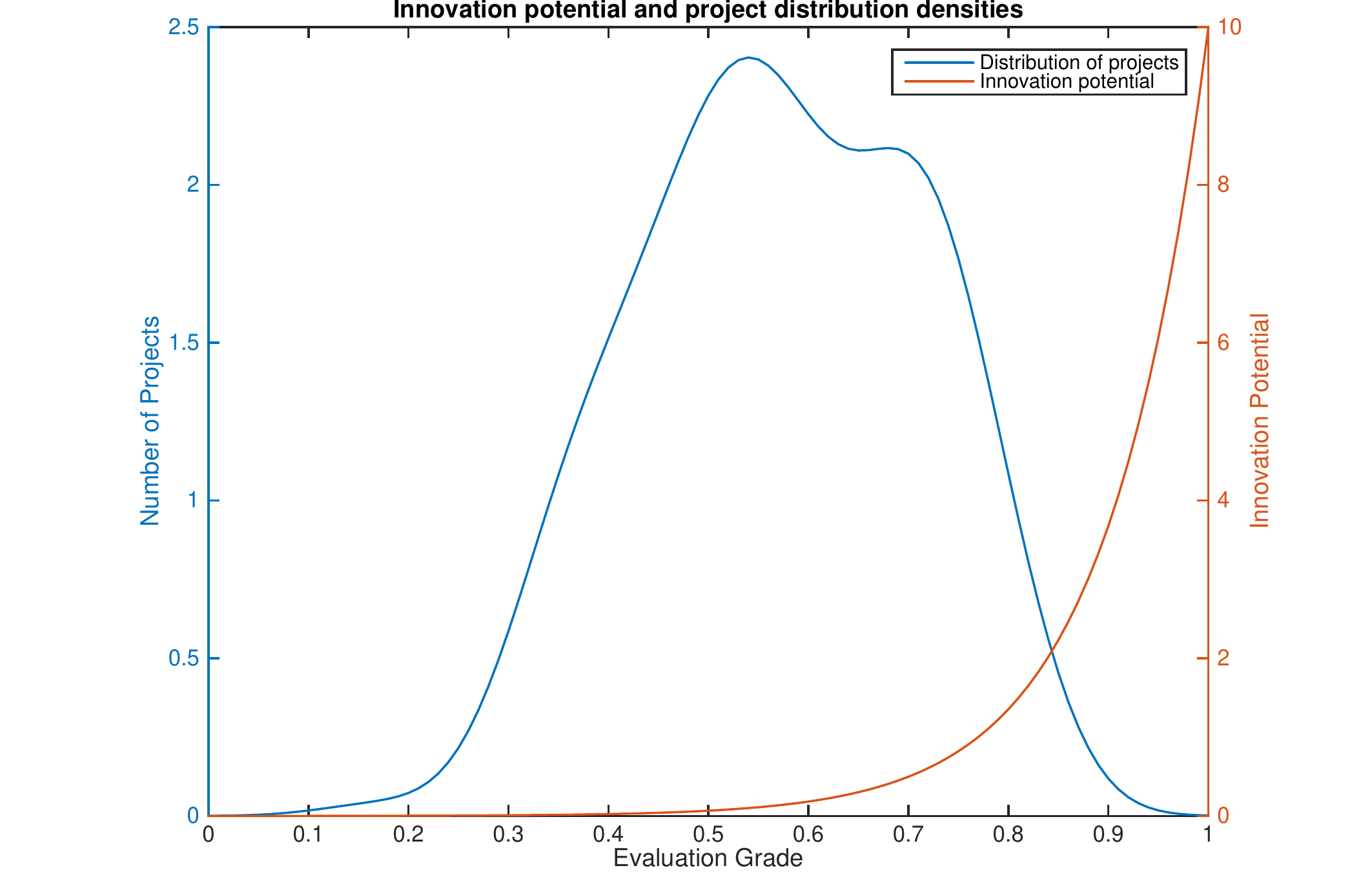}
\caption{The estimated distribution of scores from an FP7  evaluation exercise, superposed to a theoretical innovation density curve.}
\label{RealDensity}
\end{figure}

The distribution of projects with respect the score is very asymmetric. The estimated probability density obtained from real evaluation exercises (Fig.\ref{RealDensity}) show that evaluations tend to cluster at a larger value than the average grade, with very small tails at high and low levels. The data here has been obtained from past evaluation results of the FP7 EU program (Barbante, 2015, Pers. Comm.) . Few projects score the maximum and few projects are indeed so bad to deserve the minimum score, as a result the distribution has a internal peaks. 

We can also describe the innovation content of a project with  an {\em innovation density} $v(x)$ , meaning that $v(x)$ is the innovation content of a proposal with score $x$.  The total innovation of the proposals up a certain score is then given by the integral
$$
I(\lam) = \frac{1}{I_0}\int_0^\lam v(x) p(x)\, dx
$$
where a normalisation has been introduced so that the maximum  value of innovation achievable with this particular set of proposals is one.

The innovation content is  an abstract quantity that indicate the amount of new results, advancements or in general new science attained by the project. It is difficult to model such function, but it seems that in order to have the evaluation process make any sense at all it must be a growing function of the score $x$, possibly a very nonlinear function as we expect that the best project will have a considerably larger potential than the rest. In this case we can choose a simple behaviour:
$$
v(x) = \exp(\alpha x) -1
$$
the scale $\alpha$ will give us the strength of increase with the increasing score.
Fig.\ref{RealDensity}  shows an example of such a density for $\alpha =10$. 

\begin{figure}[h!]
\centering
\includegraphics[width=\columnwidth,clip,trim= 0 200 0 200]{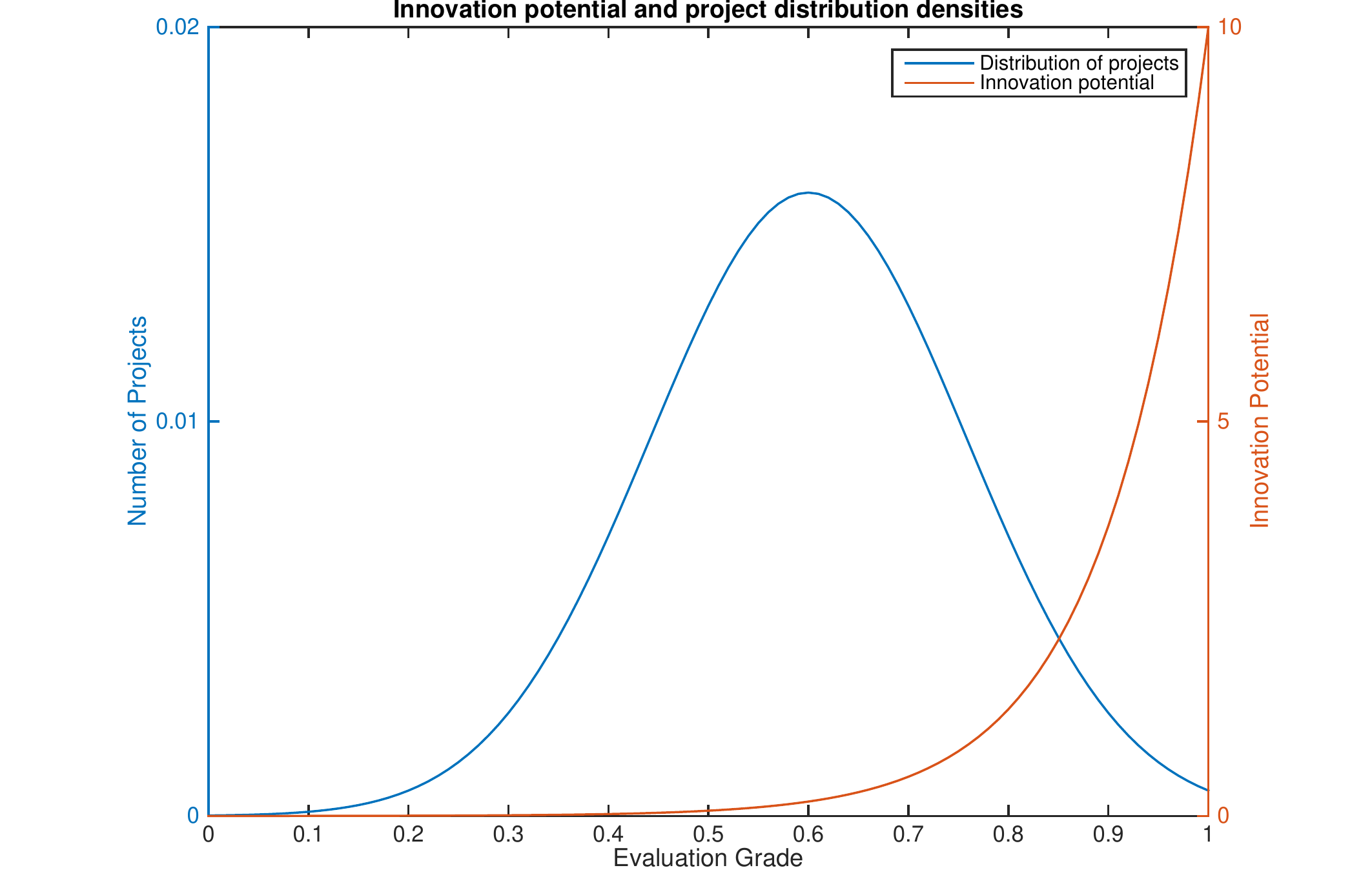}
\caption{The project density function and the innovation density functions, using $x_0 = 0.6, \sigma =0.1$ and $\alpha =10$}
\label{Density}
\end{figure}

The shape of the project density function suggests that we can model it with a simple function
$$
p(x) =  \exp \left( - \frac{(x-x_0)^2}{\sigma}\right)
$$
where $x_0$ is the score at the peak of the distribution and $\sigma$ is the width of the distribution. Fig.\ref{Density} shows two examples together of the project density function and of the innovation density function.

\section{The optimal choice}

A typical procedure would proceed to accept proposals starting from the maximum score working down the list toward lower values, usually until funds are exhausted. The issue we would like to investigate is if there is a way to determine a theoretical optimal choice (in some sense) to choose the funding threshold , $\lam$, such that proposals scoring higher than that will be retained and the others declined.  We can define the problem as follows: we need to find the cut $\lam$ that gives the maximum total innovation $I(\lam)$ with the minimum number of proposals. The total innovation for the proposal retained above the cut $\lam$ is therefore given by
$$
J_1(\lam) = \int_\lam^1 p(x) v(x) \, dx
$$
and the number of retained proposal is
$$
N_R(\lam)=\frac{1}{N_1}\int_\lam^1 p(x) \, dx
$$
we would like to get the maximum innovation with the minimum of proposals, so the desired threshold $\lam$ is such that $max(J_1)$ and $min(N_R)$. We can observe however that the minimum of retained proposal is equivalent to maximising the number of declined proposals, so we can use the total number  of declined proposals

$$
J_2(\lam) = \frac{1}{N_1}\int_1^\lam p(x) \, dx
$$
such that
$$
\min(N_R)=\min \left( \frac{1}{N_1}\int_\lam^1 p(x) \, dx \right) = \max  \left( \frac{1}{N_1}\int_0^\lam p(x) \, dx \right)= \max(J_2)
$$
and since $J_1$ and $J_2$ are positive,  we can look for the maximum of a cost function $J$
$$
J(\lam) = \int_\lam^1 p(x) v(x) \, dx \, + \int_0^\lam p(x) \, dx
$$

\begin{figure}[h!]
\centering
\includegraphics[width=\columnwidth,clip,trim= 0 200 0 200]{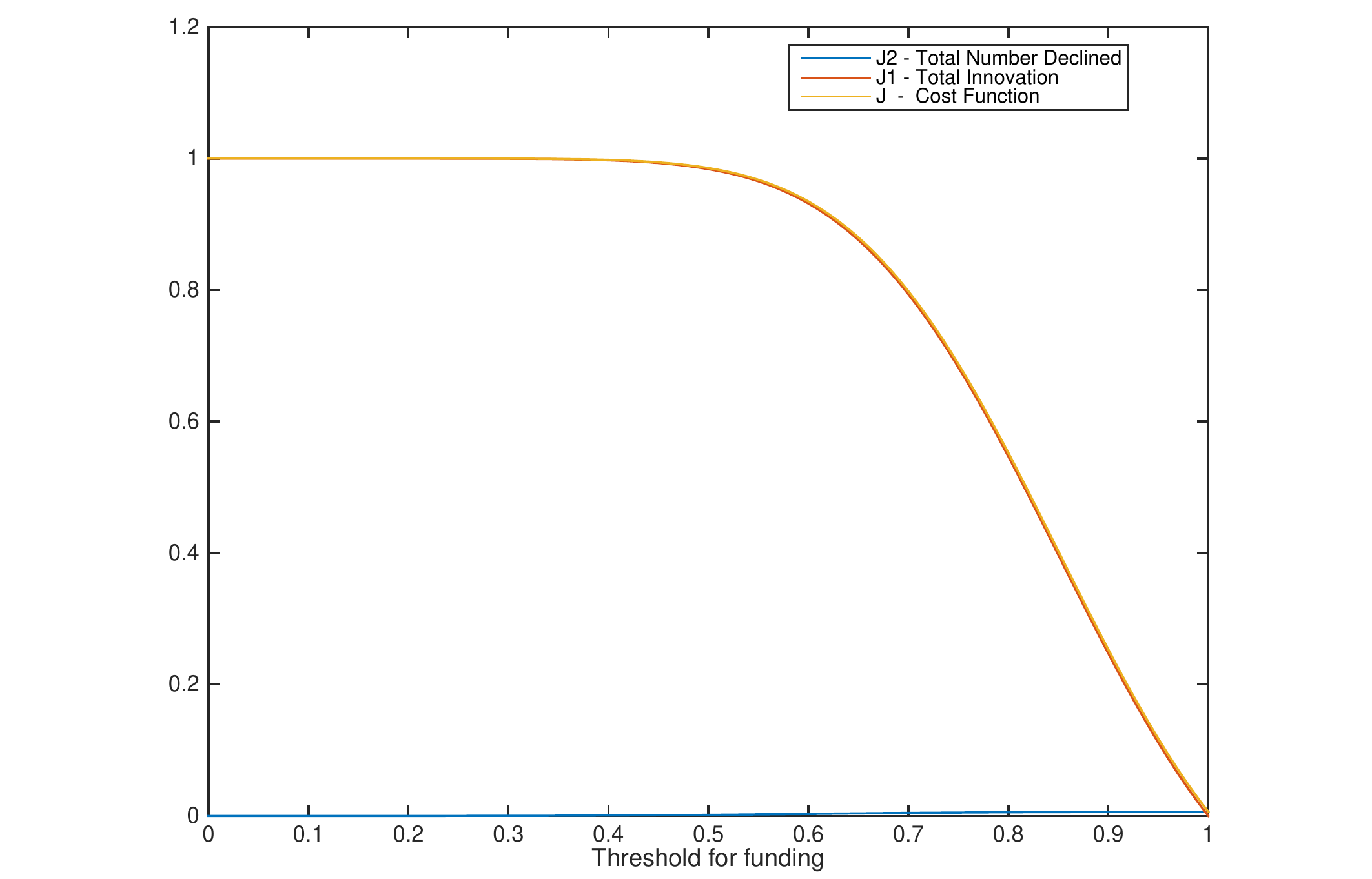}
\caption{The total number of declined proposals, $J_2$,  the total innovation achieved, $J_1$ and the cost function $J$ as a function of the threshold value $\lam$. The case shown here is   for a project density function with $x_0 = 0.6, \sigma =0.1$ and for an innovation density function with a scale $\alpha =10$.}
\label{Cost}
\end{figure}

Fig.\ref{Cost} shows the behaviour of these functions. The total innovation is  large when all projects are retained and because of the steep behaviour of the innovation density function the innovation is really all concentrated in the best projects, i.e. those scoring 0.7 and higher, as it is expected. The number of projects as a function of the threshold drops less rapidly as a consequence of the internal maximum of the density function. As a result the cost function has also an internal maximum. For the case in the picture the maximum total innovation with the minimum number of projects can be obtained with a score threshold of 0.74, corresponding to retaining 26\% of the projects.

\section{Sensitivity Tests}

\begin{figure}[h!]
\centering
\includegraphics[width=\columnwidth,clip,trim= 0 200 0 200]{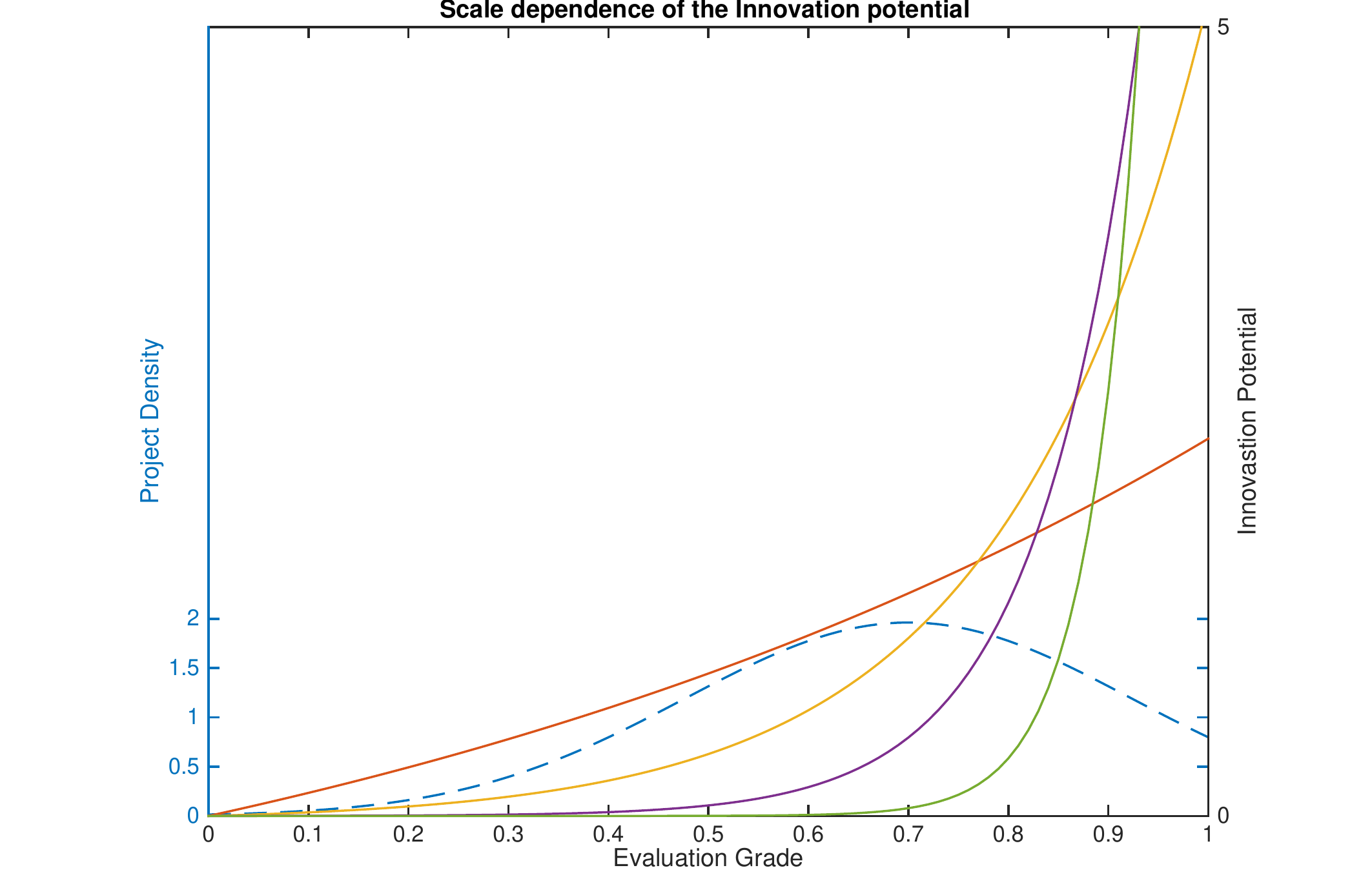}
\caption{Innovation density functions for $\alpha = 1,5,10,20$. A project density function (dashed line) using $x_0 = 0.6, \sigma =0.1$ is also shown for reference}
\label{Innovation}
\end{figure}

Different behaviour of the innovation functions will fix the maximum innovation obtainable from a given set of proposals, because the innovation is basically the overlap integral between the projects distribution and the innovation curve. Fig.\ref{Innovation} shows the innovation for various values of the scale factor. As the innovation gets more concentrated towards the higher values of the score fewer and fewer project will contribute to the total innovation. Conversely, assuming a weaker dependence of the innovation on the score requires more projects to contribute to the totale innovation.

\begin{figure}[h!]
\centering
\includegraphics[width=\columnwidth,clip,trim= 0 200 0 200]{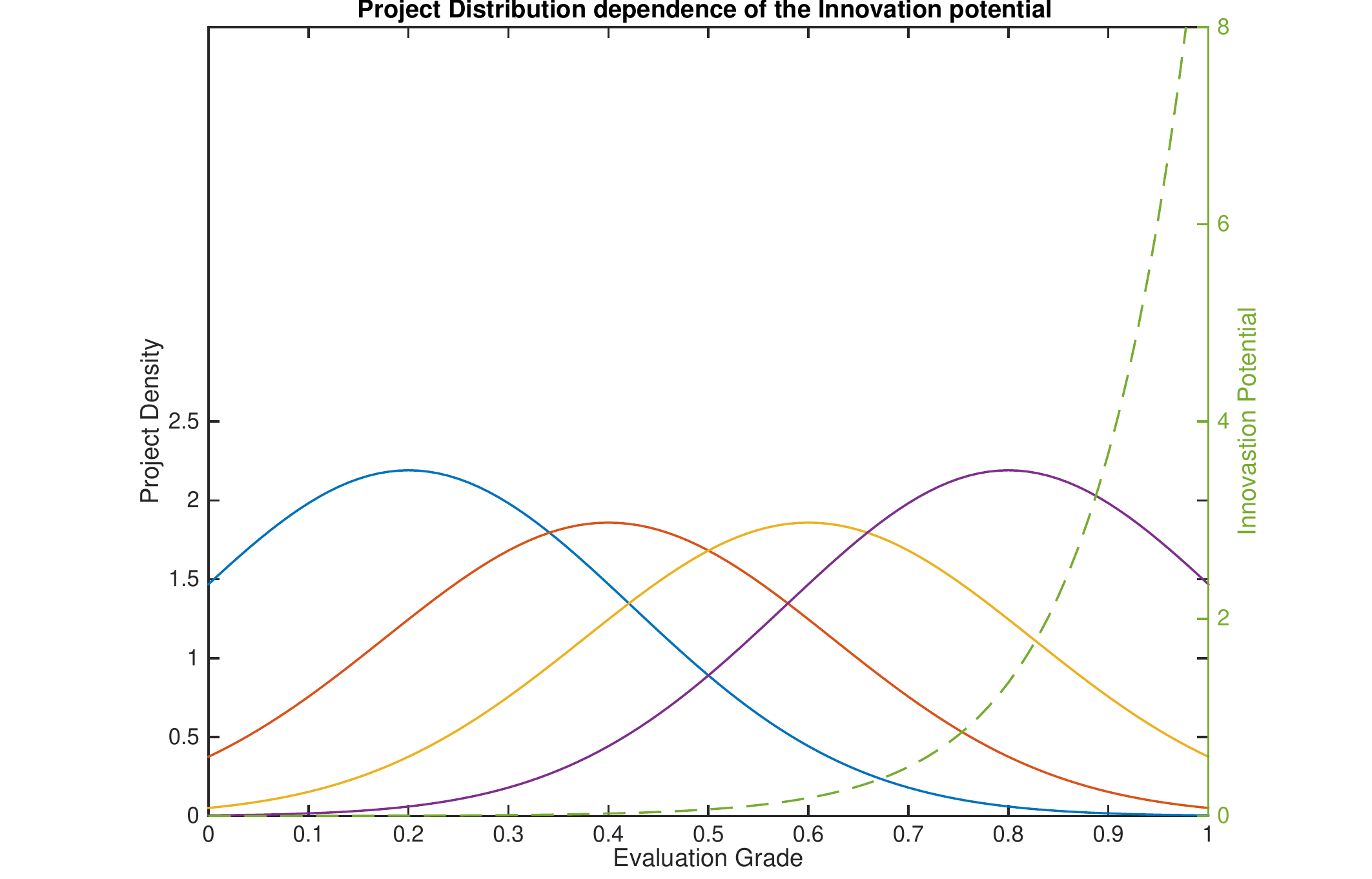}
\caption{Project density functions using $x_0 = 0.2, 0.4, 0.6, 0.8$, $\sigma =0.1$. An innovation density with $\alpha =10$ (dashed line) is also shown for reference}
\label{Quality}
\end{figure}

By there same token, Fig.\ref{Quality} shows what happened for various project density functions. If the project density distribution is peaked at low values there will be a limited overlap and a small total  
innovation is generated, whereas a distribution peaked toward high values will result in a large total innovation.

We can get higher thresholds if we assume a faster increase of the innovation density with the score. Table  \ref{ScaleTable} shows some results with different scales in the innovation function. The scale gives a measure of the gap between the best and the worst proposal. It is  interesting to note that it is very difficult to get  large rejection values, above 95\%, corresponding to single digits success rates. This is quite reasonable for innovation that differs by orders of magnitude, but it is sub-optimal if there is a more uniform innovation values of the projects. It appears that for more moderate innovation densities values around 30\% are more reasonable. Interestingly, these numbers corresponds to experimentally obtained values from the National Science Foundation \citep{NSF2014}. In the period 2004-2013  the overall success rates for proposals, defined as the ratio between the proposal funded with respect to the total number of proposal presented, has always been higher than $20\%$, peaking at $32\%$ in 2009, decreasing to  $22-24\%$ in the following years. Disciplinary differences  can be seen, GeoSciences peaked in 2009 at $45\%$, in the same year Social and Behavioural and Economics sciences  reached $30\%$ and Mathematics and Physics $40\%$. In Europe, the FP7 cooperation program yields a general success rate of $17\%$, averaged over the five years of the program and the environmental program in particular yielded a similar result \citep{Fp7}. Statistics for the grants from the European Research Council are much lower. Success rates for Starting Grant from 2007 to 2014 were around $10\%$, Advanced Grant fared a little better around $13\%$ (ERC Webpage, 2014).

\begin{table}[]
\begin{center}\begin{tabular}{|c||c|c|c|}\hline Center & {\em Threshold}& {\em Retained Projects}& {\em  Innovation Obtained}
 \\\hline 0.2 & 46\% & 15\% & 0.85\\ 
 \hline 0.4 & 61\% & 17\% & 0.85\\ 
 \hline 0.6 & 74\%& 23\% & 0.83\\ 
 \hline 0.8 & 83\% & 32\% &0.80\\ 
 \hline \end{tabular} \caption{Optimal thresholds, project retained and normalised total innovation achieved by retained projects for various values of the position of the center of the project density function. The innovation scale used here has $\alpha =10$. A distribution shifted very much toward low values indicates a set of proposal projects of low innovation potential therefore only a small fraction are worth retaining to achieve the nest result. In this case the total (unnormalized) innovation value is much less.
 \label{ScaleTable}}
\end{center}

\end{table}

\begin{table}[]
\begin{center}\begin{tabular}{|c||c|c|c|}\hline Scale $\alpha$ & {\em Threshold}& {\em Retained Projects}& {\em  Innovation Obtained}
 \\\hline 1 & 68\% & 49\% & 0.64\\ 
 \hline 5 & 74\% & 37\% & 0.71\\ 
 \hline 10 & 79\%& 27\% & 0.82\\ 
 \hline 20 & 85\% & 17\% & 0.91\\ 
 \hline \end{tabular} \caption{Optimal thresholds, project retained and normalised total innovation achieved by retained projects for various values of the scale of the innovation density function. The project density function has been kept fixed at $x_0=0.7, \sigma =0.1$. A steeper behaviour od the innovation function results in fewer projects retained and higher innovation. 
 \label{ScaleTable}}
\end{center}

\end{table}

\section{Conclusion}

This simple model indicates that the selection process will yield optimal impacts only if the realistic distribution of the innovation potential is considered. The example from a realistic evaluation shows that the innovation potential, as it is measured by the score, is distributed  into a larger portion of initiatives and ideas -- i.e. proposals -- that simply those scoring at the maximum level. As a consequence, optimal innovation can be reached only by accepting a wider range of proposals. Imposing very high thresholds, like in the ERC case, will have a sub-optimal impact.

The model is a very simplified analysis that is of course missing cost considerations and more generally policy constraints and decisions that can affect the general mechanisms outline here, but the main issue of having a mismatch of the value of the proposals and the implicit innovation potential looks fundamental. Unless, of course, we admit that the evaluation process is inaccurate and the scores do not reflect the innovation potential. Probably a scoring system that is not expressed in terms of continuous values, but is designed with categories as Certainly Reject, Certainly Accept, Accept with Reserve, for instance, would be less mechanistic and more in line with the ultimate policy goal to maximise innovation and production of knowledge. It is clear that a system with very low values of acceptance resembles a lottery system and though it is right to have an award system in place to recognize exceptional achievements, it is doubtful it can be an effective instrument of policy. Nobel prizes can come after the research is done, but they are not the way to stimulate and encourage new research.

\begin{acknowledgment} 
The support of the GEMINA project funded by the Ministry for University and Research and by the Ministry of Environment, Land and Sea of Italy is gratefully acknowledged. The National Center for Atmospheric Research is supported by the U. S. National Science Foundation.
\end{acknowledgment}


\bibliographystyle{ametsoc}
\bibliography{project}

\end{document}